# Phonon anharmonicity in Cu-based layered thiophosphates

Rahul Rao,[1*] Michael A. Susner[1]

[1]Materials and Manufacturing Directorate, Air Force Research Laboratory, 2179 12th Street, Wright-Patterson Air Force Base, Ohio 45433, USA

**Abstract**

In this work, we performed temperature-dependent Raman spectroscopy studies on single crystals of three layered metal thiophosphates: $CuInP_2S_6$, $CuCrP_2S_6$, and $CuInP_2S_6$-$In_{4/3}P_2S_6$ heterostructures. These materials are emerging multiferroics, with $CuInP_2S_6$ and $CuInP_2S_6$-$In_{4/3}P_2S_6$ exhibiting room temperature ferrielectric polarization while $CuCrP_2S_6$ is both anti-ferroelectric and antiferromagnetic at low temperatures (<145 and <30 K, respectively). We studied two prominent peaks in the Raman spectra of these compounds, namely their phosphorus-phosphorus (P-P) and phosphorus-sulfur (P-S) stretching modes. First order linear fits to the mode softening yielded coefficients similar to other layered materials. Furthermore, analysis of the phonon decay mechanism revealed the dominant role of anharmonic effects rather than thermal expansion. In particular, the decay of the optical phonons was found to be governed by three-phonon processes in $CuInP_2S_6$ and $CuCrP_2S_6$ and by the four-phonon process in the $CuInP_2S_6$-$In_{4/3}P_2S_6$ heterostructures. These phonon lifetimes were found to be less than 1 ps, indicating high scattering rates and highlighting the high degrees of anharmonicity in these materials.

## 1. INTRODUCTION

Layered metal thiophosphates (MTPs) such as $CuInP_2S_6$ and $CuCrP_2S_6$ are an emerging class of functional van der Waals materials that exhibit correlated electron properties including room-temperature ferroelectricity and tunable quadruple-well polarization [1,2], field-induced ferromagnetic states [3], and superconductivity at high pressures [4]. Interestingly, many of these compounds exhibit correlated electron properties down to the few- and mono-layer limits, making them excellent candidates for nanoelectronics at this scale [5,6]. In the MTPs, the ferroic properties result from the lamellar structure intrinsic to these compounds; an example of the lamellar structure of $CuInP_2S_6$ is shown in Fig. 1a. Within each S-terminated lamella are three octahedral sites bounded by S atoms; one of these is reserved for P-P dimers which contribute to the structure of the $[P_2S_6]^{4-}$ anion sublattice. The remaining two sites act as hosts for metal cations which in turn can carry magnetization or electric polarization into the compound. The structure-dependent properties and phases in MTPs are tunable; varying temperature and pressure through heating/cooling and hydrostatic compression, respectively, causes changes in the atomic/electronic structure which in turn alters the correlated properties observed in these materials.

---

* Correspondence – rahul.rao.2@us.af.mil

While most of the focus on the layered MTPs has been devoted to the measurement of their electromagnetic and ferroelectric properties, less attention has been paid to phonon transport within these materials. A high degree of optical phonon anharmonicity is believed to be responsible for the highly anisotropic elastic properties in $CuInP_2S_6$ [7]. Moreover, the recent observation of high and low polarization states in $CuInP_2S_6$ also revealed that the low polarization state is accompanied by a large negative out-of-plane piezoelectric coefficient ($d_{33}$) [2]. This was attributed to the high degree of anharmonicity in the quadruple-well potential as a function of Cu displacements. Anharmonic effects are also present in other 2D MTPs - recent theoretical calculations revealed that a strong anharmonic coupling between a polar mode and a symmetric longitudinal optical phonon mode is responsible for stabilizing ferrielectricity in $CuInP_2Se_6$ [8]. For both $CuInP_2S_6$ and $CuInP_2Se_6$, it was suggested that lattice strain could be used to tune the phonon anharmonicity and thereby their polarization states [2,8]. Finally, high degrees of lattice anharmonicity imply an increase in phonon-phonon interactions, resulting in high phonon scattering rates, which adversely affect thermal transport within the materials. Thus, the low out-of-plane and in-plane thermal conductivity of $CuInP_2S_6$ (~0.3 and ~2 $Wm^{-1}K^{-1}$ at room temperature, respectively) [9] are likely due to strong anharmonic effects [10] and suggest that $CuInP_2S_6$ (and potentially other 2D MTPs) could be interesting choices as thermal insulating materials or even thermoelectric materials, provided other properties (Seebeck coefficient and electrical conductivity) are optimized accordingly.

An effective way to probe phonon-phonon interactions in materials is through temperature-dependent Raman spectroscopy. In the past, such studies have revealed insights into the phonon decay mechanisms in elemental 2D materials such as graphene and black phosphorus [11], as well as several 2D transition metal chalcogenides such as $MoS_2$ [12], $WS_2$ [13,14], $WSe_2$ [15], $MoSe_2$ [16], $MoTe_2$ [17], $ReSe_2$ [18], and GaSe [19]. Here we present a study anharmonic phonon interactions using temperature-dependent Raman study in three layered MTPs: $CuInP_2S_6$, $CuCrP_2S_6$ and $CuInP_2S_6$-$In_{4/3}P_2S_6$ heterostructures. $CuInP_2S_6$ (CIPS) is a room-temperature ferroelectric with a high Curie temperature ($T_C$)

~315 K. When synthesized with Cu deficiencies, the material spontaneously segregates into ferrielectric CuInP$_2$S$_6$ and paraelectric In$_{4/3}$P$_2$S$_6$ (IPS). The IPS domains exert a chemical pressure over the CIPS domains, and the interplay between the two sub-lattices leads to an increase in the overall $T_C$ to ~335 K for highly Cu-deficient CIPS [1,20]. CuCrP$_2$S$_6$ (CCPS) is a multiferroic material that is paraelectric at room temperature and transitions to an antiferroelectric phase at low temperatures ($T_C$ ~145 K). At even lower temperatures (~30 K), the material exhibits antiferromagnetic ordering. As mentioned above, the structures of all three MTPs consist of S$_6$ octahedra with metal cations (Cu, In or Cr) and P-P pairs confined within them. The S$_6$ octahedra and the P atoms together form a structural backbone comprised of [P$_2$S$_6$]$^{4-}$ ethane-like anion groups that ionically pair with hexagonally arranged Cu and In cations in CIPS and CIPS-IPS, and with Cu and Cr cations in CCPS. A temperature-dependent redistribution of the Cu cations within the lamellae is primarily responsible for the observed (anti)ferroelectric behavior in these materials [21].

All three MTPs exhibit Raman spectra with several vibrational modes – low-frequency cation and anion libration and deformation modes as well as anion (P-P and P-S) stretching modes [22–25]. In this paper, we report temperature-dependent (between 77 and 500 K) Raman studies on CIPS, CIPS-IPS and CCPS, particularly focusing on the behavior of the P-P and P-S modes, which are out-of-plane stretching modes with atomic displacements predominantly perpendicular to the stacking direction of the layers. Thus, a study of the decay mechanisms of these vibrational modes would give us insights into the phonon-phonon interactions these modes are expected to Upon heating, we observed anharmonic softening and broadening of these modes. Modeling of the frequencies and linewidths revealed a much stronger contribution of lattice anharmonicity compared to thermal expansion in the decay of the optical phonons into lower energy phonons, with the predominant decay processes occurring by three-phonon decay in CIPS and CCPS, and by four-phonon decay in CIPS-IPS.

## 2. MATERIALS AND METHODS
### 2.1 CRYSTAL SYNTHESIS

Single crystals of CCPS, CIPS and CIPS-IPS were grown by vapor transport techniques as detailed in previous publications (*cf*. Refs. [1,5,25,26]). For synthesis of CCPS, Cr powder was first reduced under forming gas (95% Ar, 5% H$_2$) to remove any oxide contamination. The precursors were then weighed to achieve a molar ratio of Cu:Cr:P:S of 1:1:2:6, with ~5 % excess P included. The materials were inserted into a thick-walled quartz ampoule that was sealed under vacuum together with ~100 mg of I$_2$ to act as a vapor transport agent and placed in a tube furnace. The temperature was slowly ramped to 300 °C over a period of 10 h, after which point it was held for a further 10 h. After this initial heating step, the furnace was heated to 700 °C over a period of 13 h and held for ~7 days, followed by cooling down to room temperature over a period of 35 h. For CIPS and CIPS-IPS, we first synthesized $In_2S_3$ from elements (Alfa Aesar Puratronic, 99.999% purity, sealed in an evacuated fused silica ampoule and reacted at 950 °C for 48 hours) and subsequently reacted it with the necessary quantities of Cu, P and S (all from Alfa Aesar Puratronic) to obtain the pure phase $CuInP_2S_6$ as well as a Cu-deficient composition $Cu_{0.4}In_{1.2}P_2S_6$. The starting materials were sealed in a fused silica ampoule with ~80 mg of $I_2$ and loaded into a tube furnace. The furnace was slowly ramped to 775°C over a period of 24 hours and held at that temperature for 100 hours. Afterwards, the samples were cooled at a rate of 20°C/hr. to ensure the growth of large domains.

## 2.2 RAMAN SPECTROSCOPY MEASUREMENTS

Temperature-dependent Raman spectroscopy measurements were conducted in a Renishaw inVia Raman microscope. The incident excitation (785 nm, ~2 µm spot size) was directed on to the sample through a Coherent THz-micro low-frequency module, and coupled to the Raman microscope with a fiber optic cable. A 50x magnification long-working distance objective lens was used to focus on to freshly cleaved crystals mounted within a temperature stage (Microptik). By measuring with a laser power meter, the laser power at the sample was set to ~ 1 mW to avoid additional heating from the laser. Raman spectra were collected in a nitrogen ambient at temperatures ranging between 77 – 550 K. Spectra were collected at 5 - 25 °C temperature increments with 15 s acquisition times and 4 accumulations. Spectral analysis was performed (in Igor Pro) by cubic spline baseline subtraction and Lorentzian peak fitting to extract frequencies and linewidths.

## 3. RESULTS AND DISCUSSION

Room-temperature Raman spectra from CIPS, CIPS-IPS and CCPS crystals are shown in Figure 1b. Each spectrum was fitted to Lorentzian peaks (plotted below the experimental spectra), with the overall fits overlaid on top of the data. As can be seen in Fig. 1, spectra from all three materials exhibit similar peaks that can be grouped into four frequency ranges. Peaks below 100 $cm^{-1}$ correspond to extended librations and translations involving the metal cations, while peaks between 150 – 200 $cm^{-1}$ and 200 – 350 $cm^{-1}$ correspond to deformations of the S-P-P and S-P-S bonds within the octahedra, respectively. The sharp peaks around 380 cm$^{-1}$ primarily involve out-of-plane P-P stretching vibrations [ν(PP)], and the high-frequency modes between 550 and 600 cm$^{-1}$ are due to out-of-plane P-S (or PS$_3$) stretching vibrations [ν(PS)].[22,24,25] Among the three MTPs, CIPS-IPS exhibits the highest number of Raman peaks due to contributions from the CIPS and IPS sub-lattices within the self-assembled CIPS-IPS heterostructures [24,27].

Temperature-dependent spectra were collected between 77 – 500 K and are displayed in heat maps (2D maps with the color scale corresponding to spectral intensities) in Figs. 2a – 2c with the corresponding Curie temperatures indicated by the dashed horizontal lines. Waterfall plots of these spectra are included in the supplementary material, Figs. S1-S3. The transition to the paraelectric state and the loss of polarization in these MTPs is driven by the movement of the Cu$^+$ ions. At room temperature, the Cu$^+$ cations in CIPS and CIPS-IPS primarily occupy the upper sites within the S$_6$ octahedra (within the ferrielectric phase for the latter case). Upon heating, they redistribute within the lamellae between the upper and lower positions and loss of polarization occurs when the Cu$^+$ ion occupancy becomes equal between the tops and bottoms of the octahedra [21]. The Cu$^+$ ion redistribution in the three MTPs is accompanied by a number of significant changes in the lattice vibrational modes such as the disappearance of peaks as well as discontinuities in peak frequencies, widths, and intensities. Some of these changes can be observed in the vicinity of the Curie temperatures (dashed lines) in Figs. 2a-2c, especially for peaks below 350 cm$^{-1}$, *i.e.* vibrational modes corresponding to anion and cation deformations. There are significant changes in the intensities of some of these modes. Detailed discussions on the effect of the phase transitions on their frequencies, linewidths and intensities, in CIPS, CIPS-IPS, and CCPS have been described elsewhere [22,24,25]. However, some vibrational modes, such as the ν(PP) and the ν(PS) modes (and labeled in Fig. 2a) persist throughout our studied temperature range and do not exhibit significant discontinuities. The P-P and P-S stretching vibrations are parallel to the stacking direction of the lamellae, making them especially sensitive to temperature-induced changes just like the out-of-plane vibrational modes in other 2D materials [14]. Below we focus on an analysis of the temperature dependence of these modes in the three MTPs.

Figures 3a and 3b plot the frequencies of the ν(PP) and the ν(PS) modes against temperature. We see a clear redshift in mode frequencies upon heating, which can be attributed to anharmonic effects [28]. For most materials, Raman phonon modes exhibit a linear temperature dependence in the intermediate temperature range (200 – 400 K) [29] and the temperature-dependent mode frequency $\omega_i(T)$ can be fit to the following relation:

$$\omega_i(T) = \omega_i^0 + \chi_i T. \tag{1}$$

Here $\omega_i^0$ is the extrapolated frequency at 0 K and $\chi_i$ is the first order temperature coefficient for the $i^{th}$ vibrational mode. The coefficient values obtained from the fits are listed in Table 1. For the ν(PP) mode (Fig. 3a), we observed similar values for CCPS and CIPS (-0.0125 and -0.0127 cm$^{-1}$K$^{-1}$, respectively). The coefficient for CIPS-IPS was slightly lower (-0.0107), implying a slightly weaker dependence of the ν(PP) mode frequency on temperature in CIPS-IPS. Overall, the first order coefficient values are similar to those observed in other layered materials such as transition metal dichalcogenides [12,16]. The first order coefficients for the ν(PS) mode (Fig. 3b) were larger than those of the ν(PP) mode, ranging from -0.0236 to -0.0376 cm$^{-1}$K$^{-1}$. The higher values indicate a stronger dependence of the P-S stretching mode on temperature compared to the P-P mode. Within a lamella the P-P bonds are aligned along the stacking direction and surrounded by S atoms in the S$_6$ octahedra. Upon heating, the increasing movement and redistribution of the Cu$^+$ ions within the octahedra results in significant distortions and strains in the P-S bonds, as discussed in detail in our recent Raman spectroscopy and X-ray diffraction study [24]. The different values of $\chi_i$ for the ν(PS) mode between the three MTPs hints at varying degrees of structural distortions, with the highest occurring in CCPS ($\chi$ = -0.0376 cm$^{-1}$K$^{-1}$).

While we fit straight lines to the temperature dependence of the ν(PP) and ν(PS) modes as shown in Figs. 3a and 3b, their dependence on temperature is in fact non-linear and a true description of the frequency (and linewidth) dependence on temperature must include the non-linearity. The temperature dependences of the frequencies and linewidths (full-width at half maximum intensity) of the ν(PP) modes in CIPS, CIPS-IPS and CCPS are plotted in Fig. 4a and 4b. The corresponding Curie temperatures for the materials are shown as dotted vertical lines in Fig. 4a. As mentioned above and seen in Fig. 3, the three MTPs exhibited non-linear redshifts in peak frequencies. In addition, we observe broadening of the ν(PP) modes with increasing temperature. According to the anharmonic theory of crystals, the Raman mode softening and broadening upon heating are due to quasi-harmonic thermal expansion and anharmonic phonon-phonon interactions [28,30]. Taking these into account, the temperature dependence of the $j^{th}$ phonon mode frequency $\omega(T)$ can be expressed according to the relation

$$\omega_j(T) = \omega_j^0 + \Delta\omega_j^V(T) + \Delta\omega_j^{anh}(T), \tag{2}$$

where $\omega_j^0$ is the temperature-independent harmonic frequency, $\Delta\omega_j^V(T)$ is the frequency shift due to lattice thermal expansion and $\Delta\omega_j^{anh}(T)$ is the frequency shift due to anharmonic phonon-phonon coupling [30,31]. The thermal expansion contribution to the phonon mode frequency is given by

$$\Delta\omega_j^V(T) = \omega_j^0 \left[\exp\left\{-\gamma_j \int_0^T \beta(T)dT\right\} - 1\right], \qquad (3)$$

where $\gamma_j$ is the temperature-dependent mode Grüneisen parameter and $\beta(T)$ is the volume thermal expansion coefficient. The optical phonons involved in the Raman scattering process undergo anharmonic decay into lower energy phonons; the model for this decay was first proposed by Klemens [32] and later extended by Balkanski [31] wherein an optical phonon decays into either two or three phonons by so-called three- or four-phonon decay processes. Here, for simplicity, we assumed that the optical phonon decays into two or three phonons with equal energies. We believe this is a reasonable assumption considering the high density of available states at lower energies in phonon density of states in MTP materials [33]. The anharmonic phonon-phonon interaction contribution to the frequency shift is therefore of the form

$$\Delta\omega_j^{anh}(T) = A\left[1 + \frac{2}{e^x-1}\right] + B\left[1 + \frac{3}{e^y-1} + \frac{3}{(e^y-1)^2}\right], \qquad (4)$$

where $x = \hbar\omega_j/2k_B T$, $y = \hbar\omega_j/3k_B T$, $\hbar$ is the reduced Planck constant, $k_B$ is the Boltzmann constant and $A$ and $B$ are anharmonic coefficients for the decay of an optical phonon $\omega_j$ by three- and four-phonon decay processes, respectively.

While the temperature-dependent Grüneisen parameter for the ν(PP) modes were unavailable, as a first approximation we obtained the isothermal Grüneisen parameter for the ν(PP) mode in CIPS using the Belomestnykh–Tesleva relation $\gamma \approx \frac{3}{2}\left(\frac{1+\nu}{2-3\nu}\right)$ [34,35] where ν is the Poisson's ratio (-0.06) for CIPS [36]. This relation is an approximation to the Leont'ev formula [37], which is a good estimation of γ for crystalline solids within a ~10% uncertainty. Using this relation, we get the Grüneisen parameter for the ν(PP) mode as 0.729. The temperature-dependent volume thermal expansion coefficients $\beta(T)$ were obtained from Ref. [38]. Using the values for γ and $\beta(T)$ in Eq. 3, we obtained the contribution of thermal expansion to the frequency of the ν(PP) mode (pink dashed line, lower panel in Fig. 4a). The dip in the data around 315 K reflects an increase in lattice thermal expansion across the ferrielectric-paraelectric phase transition [38]. Overall, the thermal expansion contribution was a poor fit to the experimental data, especially at temperatures above 200 K. This suggests that the temperature-dependence of the ν(PP) mode peak frequency in CIPS is dominated by anharmonic phonon-phonon interactions. The mode Grüneisen parameters and thermal expansion coefficients for CCPS and CIPS-IPS were unavailable in the literature; therefore, we could not calculate the thermal expansion contributions to the ν(PP) modes in CCPS and CIPS-IPS. Nevertheless, considering the similarities in structures [5], we expect the trends to be similar to that in CIPS, and anharmonic phonon-phonon interactions to dominate the temperature-dependence of the Raman peaks in the three MTPs.

To determine the nature of the anharmonic phonon-phonon interactions (*i.e.* three- vs. four-phonon decay), the ν(PP) mode frequency was fit to Eq. 4 by adjusting the coefficients $A$ and $B$ and setting $A$ ($B$) to zero for a three-phonon (four-phonon) decay process or varying both for a combined process. These fits are shown in Fig. 4a and the coefficients are listed in Table 1. For the fit to the ν(PP) mode in CIPS, the coefficients $A$ and $B$ were -1.67 and -0.099. The significantly lower value of $B$ compared to $A$ implies that three-phonon decay is the dominant anharmonic process in the temperature dependence of the ν(PP) mode in CIPS. This can also be seen in the fitted curves (red dotted curve in Fig. 4a), where the four-phonon process fits the experimental data poorly at the lowest and highest temperatures of our study. Similar anharmonic coefficients were found for the fit to the ν(PP) mode in CCPS, with $A$ = -3.3 and $B$ = -0.095, indicating the dominance of the three-phonon decay process in CCPS. However, for the fit to the ν(PP) mode in CIPS-IPS, we see that the anharmonic coefficients are nearly equal ($A$ = -0.37 and $B$ = -0.26). This implies that, unlike in CIPS and CCPS, the four-phonon process dominates phonon decay in CIPS-IPS. Indeed, as can be seen in Fig. 4a, the three-phonon process curve (green dashed curve) fitted the experimental data poorly between (100 – 450 K). The significant difference between the decay processes in CIPS-IPS and the other two MTPs can be attributed to the inherent structural differences between them, *i.e.* due to the self-assembled heterostructures in the CIPS-IPS leading to inter-mixed CIPS and IPS sub-lattices within each lamella. This inter-mixing could also result in a higher density of states and decay channels, making four-phonon decay viable. Phonon density of states calculations could confirm this hypothesis but owing to the complex structure of CIPS-IPS, the calculations would not be straightforward and are beyond the scope of this paper.

In contrast to the softening of phonon frequencies with increasing temperature, Raman peaks broaden upon heating, resulting in a concomitant increase in linewidth [28,30,32]. The temperature-dependence of the linewidths of the ν(PP) mode in the three MTPs are shown in Fig. 4b. All three materials exhibited similar linewidths, ranging from 4-6 cm$^{-1}$ at 77 K and broadening up to 12 cm$^{-1}$ at 500 K. The phonon lifetimes can be estimated from the experimentally measured linewidths according to the energy-time uncertainty relation $\tau = \hbar/\Gamma$, where the lifetime $\tau$ is in picoseconds, $\hbar$ is the modified Planck constant (5.3 cm$^{-1}$ ps), and $\Gamma$ (cm$^{-1}$) is the linewidth. The room temperature values for the linewidths from all three MTPs was 6-7 cm$^{-1}$, giving phonon lifetimes around 0.8 ps. These lifetimes are an order of magnitude lower than those of optical phonons in other 2D materials such as MoS$_2$, WS$_2$ [39] and hBN [40], but are similar to some layered thermoelectric materials such as Bi$_2$Te$_3$ [41], PbTe [42] and SnSe [43]. The low phonon lifetimes in the MTPs imply high scattering rates and the potential for use of these materials in thermoelectrics or thermal isolation applications [43–45].

The anharmonic temperature-dependence of the linewidths was determined using a relation similar to the one used for the peak frequencies:

$$\Gamma_j^{anh}(T) = \Gamma_0 + C\left[1 + \frac{2}{e^x - 1}\right] + D\left[1 + \frac{3}{e^y - 1} + \frac{3}{(e^y - 1)^2}\right]. \quad (5)$$

In Eq. 5, $\Gamma_0$ is the temperature-independent mode harmonic linewidth, and $C$ and $D$ are the anharmonic coefficients for peak broadening from the three- and four-phonon processes, respectively [30,31]. In the case of CCPS, we only fit the linewidths between 145 – 500 K owing to the sharp increase in linewidths below the $T_C$ [25]. The fit coefficients are listed in Table 1, and exhibited similar trends, with the coefficient for the four-phonon decay process ($D$) smaller than the three-phonon process coefficient ($C$). The $C$ value for CIPS-IPS was an order of magnitude higher than the corresponding values for CIPS and CCPS, which is similar to what was observed for the frequency of the ν(PP) mode in CIPS-IPS.

Figure 4c plots the temperature-dependence of the ν(PS) mode frequencies in the three MTPs along with the fits to the phonon decay model (Eq. 4). Similar to the behavior of the ν(PP) mode, the coefficient for the three-phonon decay process was much larger than that of the four-phonon decay process for CIPS ($A$ = -8.13, $B$ = -0.01) and CCPS ($A$ = -8.81, $B$ = -1.385). The values of $A$ and $B$ for the ν(PS) mode are listed in Table 1. For CIPS-IPS the value of $A$ was -0.73, and $B$ was -1.73. Thus, the three-phonon decay process appears to be the dominant phonon decay mechanism for the ν(PS) mode in CIPS and CCPS, while the four-phonon process is the dominant phonon decay mechanism in CIPS-IPS. This was the same as what we observed with the ν(PP) mode, and can also be observed in the fitted curves in Fig. 4c. Unlike the linewidths of the ν(PP) mode, the ν(PS) mode did not broaden monotonically and exhibited discontinuities across the Curie temperatures for all three materials (supplementary material, Fig. S4). Hence, we were unable to reliably fit the linewidths of the ν(PS) mode to Eq. 5. The linewidths for the ν(PS) modes were higher than the ν(PP) modes, ranging from room temperature values of ~7 cm$^{-1}$ in CIPS and CIPS-IPS to ~22 cm$^{-1}$ in CCPS. This implies shorter lifetimes for the $\nu$(PS) phonon (~0.25 ps in CCPS) and thus higher scattering rates.

## 4. CONCLUSION

We performed a temperature-dependent Raman spectroscopy study on three MTPs: CIPS, CIPS-IPS and CCPS. Detailed analysis was performed on two vibrational modes corresponding to stretching of P-P and P-S bonds within the lamellae. Both modes exhibited softening and broadening with increasing temperature (from 77 – 500 K). The temperature-dependence of the mode frequencies was modeled firstly by a first order linear fit, which yielded coefficients similar to other 2D materials. The frequencies and linewidths of the optical phonon modes were further modeled by considering thermal expansion and anharmonic phonon-phonon interactions. In CIPS, thermal expansion was found to fit the experimental data very poorly, indicating that optical phonon decay is dominated by anharmonic phonon interactions. Modeling of the anharmonic decay of the $\nu$(PP) and $\nu$(PS) modes revealed the three-phonon process to be the dominant mechanism in CIPS and CCPS, while the four-phonon process was found to be dominant in CIPS-IPS. The optical phonon lifetimes were found to be lower than 2D materials such as TMDs and hBN, but comparable to other layered thermoelectric materials. The short lifetimes (high scattering rates) and the phonon decay mechanisms highlight the potential for these materials in thermal insulation (and potentially thermoelectric) applications, where the four-phonon process plays an important role in the anharmonic decay of optical phonons.

**ACKNOWLEDGEMENTS**

We acknowledge support through the United States Air Force Office of Scientific Research (AFOSR) LRIR 23RXCOR003 and AOARD-MOST Grant Number F4GGA21207H002.

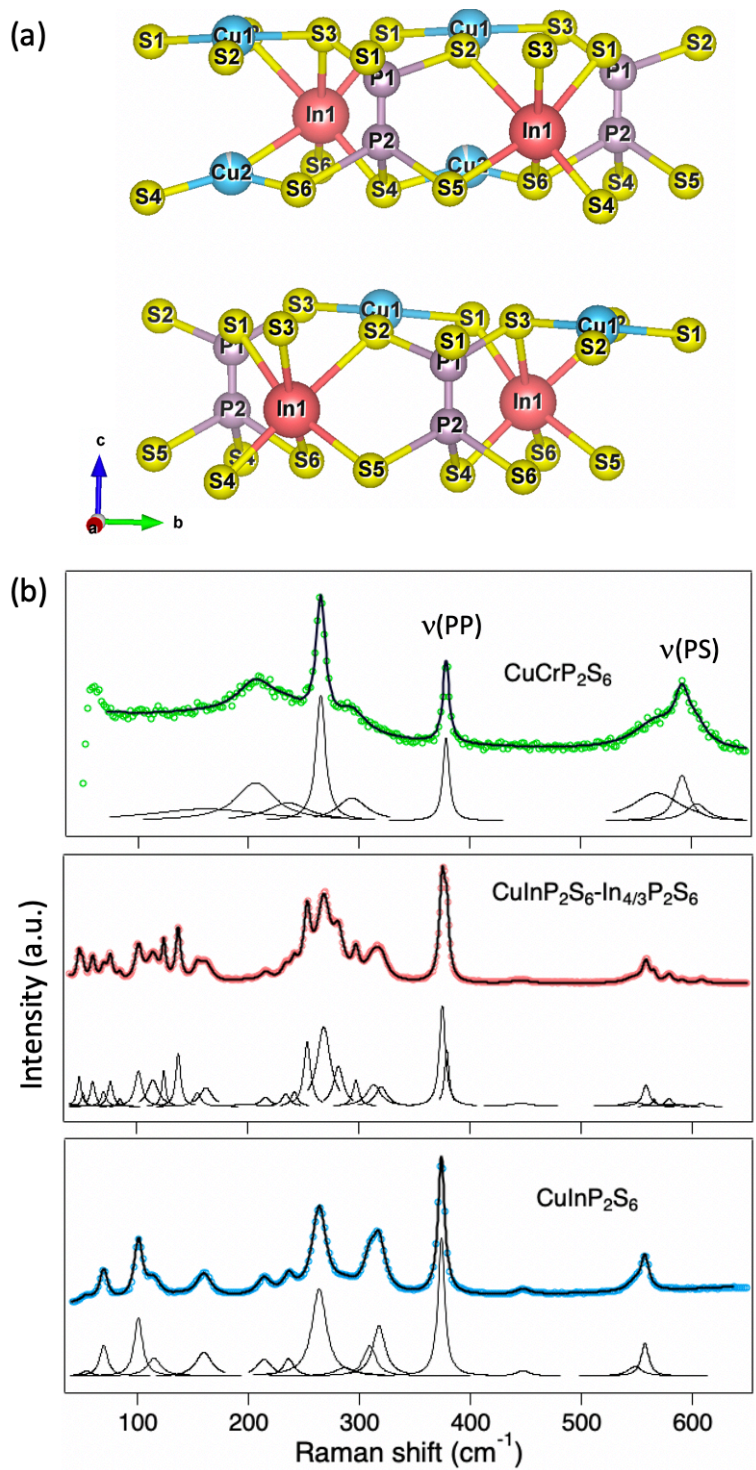

FIGURE 1. (a) Structure of $CuInP_2S_6$ in the ferrielectric phase, showing two lamellae. (b) Room-temperature Raman spectra from $CuCrP_2S_6$ (CCPS), $CuInP_2S_6$ (CIPS) and $CuInP_2S_6$-$In_{4/3}P_2S_6$ (CIPS-IPS) single crystals. The spectra have been fitted to Lorentzian peaks (shown below each spectrum) with the overall fit overlaid on top of the experimental data.

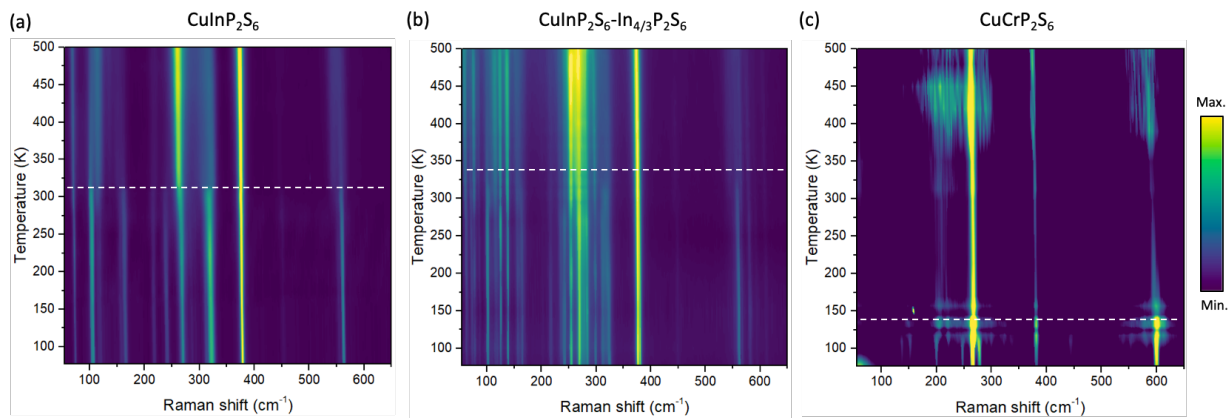

FIGURE 2. Two-dimensional heat map showing temperature-dependent Raman peak frequencies of (a) CIPS, (b) CIPS-IPS and (c) CCPS.

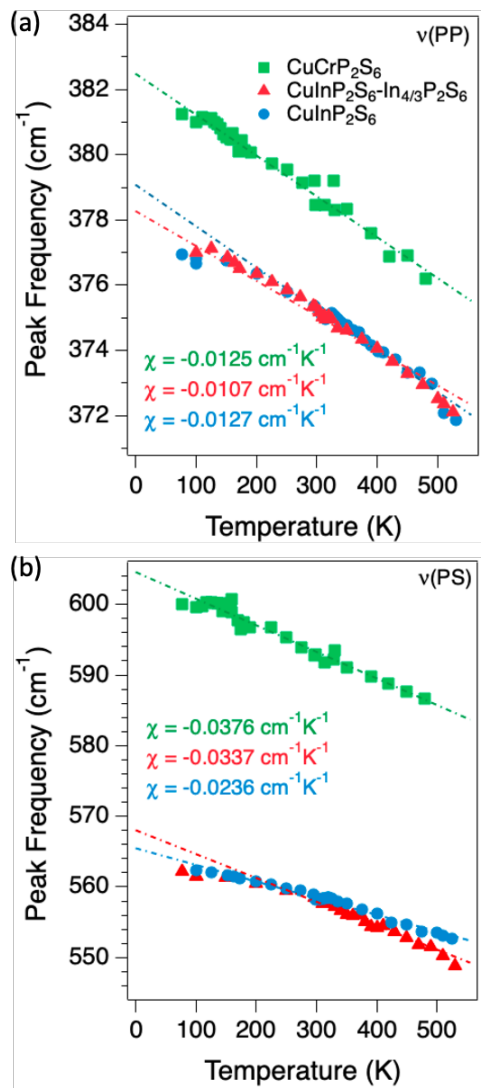

FIGURE 3. Linear fits to the (a) ν(PP) and (b) ν(PS) modes in CCPS, CIPS-IPS and CIPS. The first order temperature coefficients are listed in the insets.

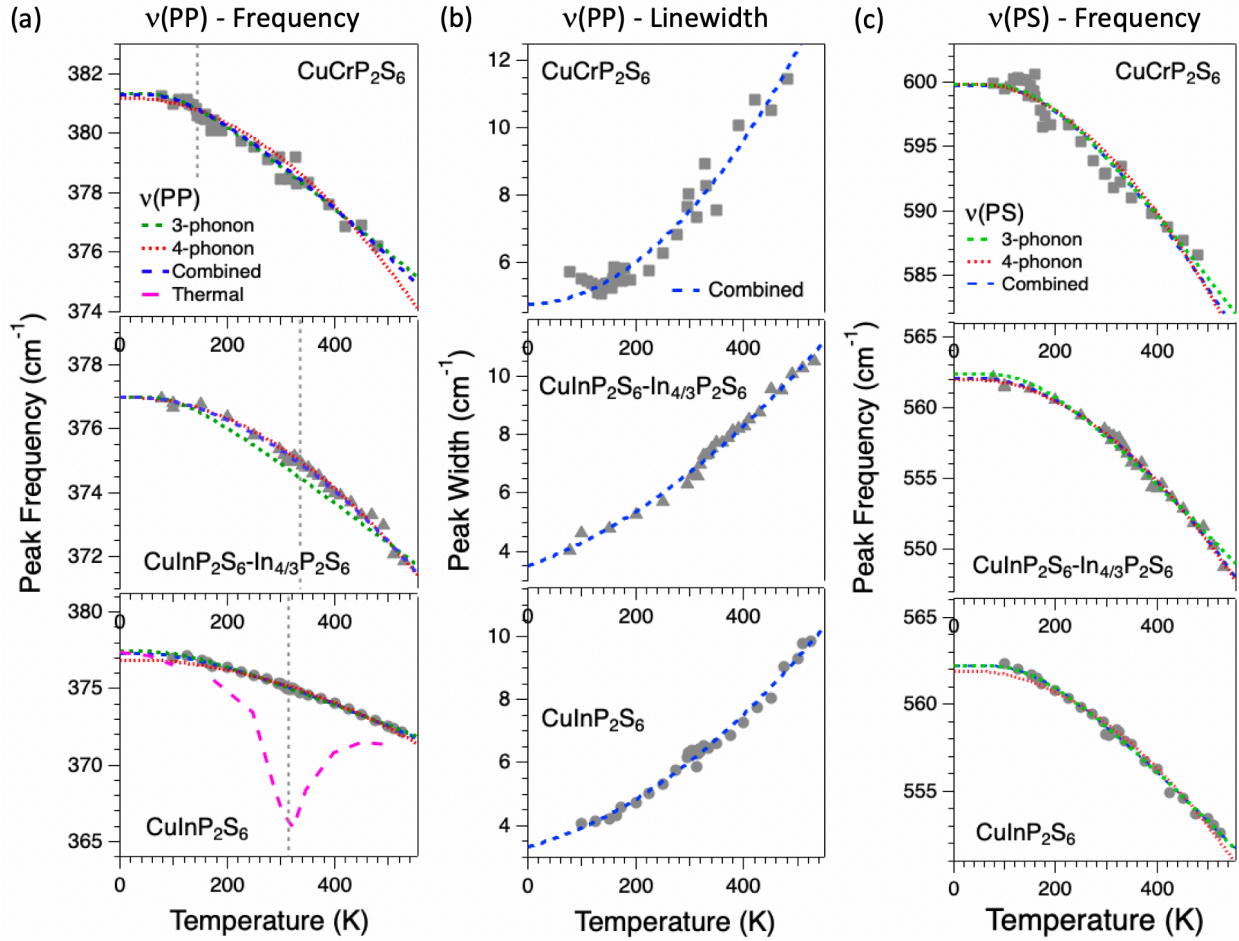

FIGURE 4. Temperature-dependent (a) frequencies and (b) widths of the ν(PP) stretching mode and (c) frequencies of the ν(PS) stretching mode in CIPS, CIPS-IPS and CCPS. The data have been fit to the phonon decay models discussed in the text.

TABLE 1. Fitting parameters for the anharmonic temperature-dependent frequencies (Eq. 4) and linewidths (Eq. 5) for the $\nu(PP)$ and $\nu(PS)$ peaks in CCPS, CIPS-IPS and CIPS, along with their extrapolated peak frequencies and widths at 0 K. Also listed are the first order temperature coefficients obtained from linear fits to the peak frequencies.

| Material | Mode | $\chi$ (cm$^{-1}$K$^{-1}$) | $\omega_0$ (cm$^{-1}$) | $A$ (cm$^{-1}$) | $B$ (cm$^{-1}$) | $\Gamma_0$ (cm$^{-1}$) | $C$ (cm$^{-1}$) | $D$ (cm$^{-1}$) |
|---|---|---|---|---|---|---|---|---|
| **CuInP$_2$S$_6$** | $\nu(PP)$ | -0.0125 ± 0.0002 | 379.05 | -1.67 | -0.099 | 3.33 | 6.9 x 10$^{-3}$ | 2.031 x 10$^{-5}$ |
| | $\nu(PS)$ | -0.0236 ± 0.001 | 570.4 | -8.15 | -0.004 | s | | |
| **CuCrP$_2$S$_6$** | $\nu(PP)$ | -0.0125 ± 0.0004 | 384.5 | -3.3 | -0.095 | 4.72 | 1 x 10$^{-3}$ | 8.129 x 10$^{-5}$ |
| | $\nu(PS)$ | -0.033 ± 0.00078 | 610.05 | -9.06 | -1.344 | | | |
| **CuInP$_2$S$_6$-In$_{4/3}$P$_2$S$_6$** | $\nu(PP)$ | -0.0107 ± 0.0005 | 377.6 | -0.37 | -0.26 | 3.51 | 1.06 x 10$^{-2}$ | 2.032 x 10$^{-5}$ |
| | $\nu(PS)$ | -0.0315 ± 0.0017 | 564.4 | -0.73 | -1.733 | | | |